\documentclass[aps,twocolumn]{revtex4-1}

\usepackage{bm}
\usepackage{amsmath,amssymb,cases}
\usepackage{braket}
\usepackage[dvipdfmx]{xcolor}
\usepackage[dvipdfmx]{graphicx}

\begin{document}

\title{Multiple-$Q$ magnetic orders in Rashba-Dresselhaus metals}

\author{Ken N. Okada}
\author{Yasuyuki Kato}
\author{Yukitoshi Motome}

\affiliation{Department of Applied Physics, University of Tokyo, Tokyo 113-8656, Japan}

\begin{abstract}
We study magnetic textures realized in noncentrosymmetric Kondo lattice models, in which localized magnetic moments weakly interact with itinerant electrons subject to Rashba and Dresselhaus spin-orbit couplings.
By virtue of state-of-the-art numerical simulations as well as variational calculations, we uncover versatile multiple-$Q$ orderings under zero magnetic field, which are found to originate in the instabilities of the Fermi surface whose spin degeneracy is lifted by the spin-orbit couplings.
In the case with equally-strong Rashba and Dresselhaus spin-orbit couplings, which is known to realize a persistent spin helix in semiconductor quantum wells, we discover a sextuple-$Q$ magnetic ordering with a checkerboard-like spatial pattern of the spin scalar chirality.  
In the presence of either Rashba or Dresselhaus spin-orbit coupling, we find out another multiple-$Q$ ordering, which is distinct from Skyrmion crystals discussed under the same symmetry.
Our results indicate that the cooperation of the spin-charge and spin-orbit couplings brings about richer magnetic textures than those studied within effective spin models.
The situations would be experimentally realized, e.g., in noncentrosymmetric heavy-fermion compounds and heterostructures of spin-orbit coupled metals and magnetic insulators.
\end{abstract}

\maketitle

\section{Introduction}\label{sec:intro}
Over a decade noncoplanar spin configurations in metals have been gathering growing interest as a source of topological transport phenomena. 
In general noncoplanarity of localized spins is characterized by the spin scalar chirality, defined as ${\mathbf S}_j\cdot{\mathbf S}_k\times{\mathbf S}_l$ for three spins spanned by sites $j$, $k$, and $l$. 
In the spin-charge coupled systems the spin scalar chirality is imprinted on conduction electrons as a fictitious magnetic field, coined as an emergent magnetic field, through the so-called Berry curvature in real space~\cite{Ohgushi2000, Nagaosa2013}.
The emergent magnetic field gives rise to a peculiar Hall effect named the topological Hall effect, distinguished from the conventional anomalous Hall effect in the presence of a ferromagnetic order. 
The topological Hall effect has been first observed in a pyrochlore magnet~\cite{Taguchi2001} and recently in metallic compounds hosting magnetic Skyrmion crystals (SkXs)~\cite{Lee2009, Neubauer2009}.

Noncoplanar spin configurations are often described by superpositions of spin helices running in different directions.
These are called multiple-$Q$ magnetic orderings. One of the latest examples is a magnetic SkX mentioned above, which can be described as a double- or triple-$Q$ ordering~\cite{Muhlbauer2009, Yu2010, Heinze2011, Nagaosa2013}.
In real space such a SkX forms a two-dimensional periodic array of spin-swirling nanometric objects, called Skyrmions. 
Each Skyrmion is characterized by a topological invariant defined by the integration of the spin scalar chirality, which guarantees the topological stability.
To date, SkXs have been experimentally identified in various noncentrosymmetric magnets, including chiral metals such as $B$20-type alloys $MX$ ($M$=Mn, Fe, Co; $X$=Si, Ge)~\cite{Muhlbauer2009,Yu2010} and $\beta$-Mn-type
Co-Zn-Mn alloys~\cite{Tokunaga2015} as well as heterostructures like a monolayer of Fe on Ir substrates~\cite{Heinze2011}.
Notably SkXs not only bring about peculiar transport of conduction electrons such as the topological Hall effect~\cite{Lee2009, Neubauer2009}, but also show their own intriguing dynamics driven by an electric current flow, resulting in current-induced motion with a remarkably-low threshold~\cite{Jonietz2010} and the Skyrmion Hall effect~\cite{Jiang2016}.
Such high mobility of Skyrmions would be potentially harnessd to future memory devices.

There are several known mechanisms for the formation of multiple-$Q$ orderings.
For SkXs observed in noncentrosymmetric 3$d$-electron systems listed above, the Dzyaloshinskii-Moriya (DM) interaction described as ${\mathbf D}\cdot{\mathbf S}_j\times{\mathbf S}_k$ plays a crucial role, which originates in the spin-orbit coupling (SOC) under broken spatial inversion symmetry.
Indeed, magnetic-field--temperature phase diagrams in those compounds including the SkX phase can be qualitatively explained by using localized spin models with ferromagnetic and DM interactions between the neighboring spins~\cite{Yi2009, Yu2010}.

On the other hand, recent theoretical studies have proposed a distinct mechanism for the formation of multiple-$Q$ orderings in centrosymmetric itinerant magnets~\cite{Akagi2012,Hayami2014,Ozawa2016,Ozawa2017prl}.
They revealed that in centrosymmetric Kondo lattice models, in which conduction electrons are coupled to localized spins, multiple-$Q$ orderings could be driven by the Fermi surface instability, irrespective of lattice types and electron fillings~\cite{Akagi2012,Hayami2014,Ozawa2016,Ozawa2017prl}.
Perturbation analyses up to fourth order with respect to the spin--charge coupling strength~\cite{Akagi2012,Hayami2014,Ozawa2016} as well as unbiased numerical simulations~\cite{Ozawa2016, Ozawa2017prl} showed that when partial nesting occurs on the Fermi surface at multiple wave vectors, in other words, when portions of the Fermi surface are connected to each other, multiple-$Q$ orders are ubiquitously favored rather than single-$Q$ orderings in the weak coupling regime.
We note that recently a SkX has been discovered in a centrosymmetric $f$-electron compound Gd$_2$PdSi$_3$, whose origin might be closely related to this mechanism~\cite{Kurumaji2018}. 

Considering the above arguments, a question naturally arises; can multiple-$Q$ orderings also show up, or if so what kind, when both the SOC and the Fermi surface instability cooperate under broken inversion symmetry?
Thus it is an intriguing task to uncover magnetic orderings in noncentrosymmetric Kondo lattice models with conduction electrons subject to the SOC. 
Recently, this issue was addressed in the case of Rashba SOC by one of the authors and his coworker, by deriving an effective spin model by the second-order perturbation with respect to the spin-charge coupling~\cite{Hayami2018}. 
Nonetheless, it would be important to solve the original noncentrosymmetric Kondo lattice model beyond the second-order perturbation, when considering the fact that in the Kondo lattice model without the SOC higher-order contributions may stabilize distinct magnetic textures from those in the perturbative regime~\cite{Akagi2012,Hayami2014,Ozawa2016}.
Moreover, it would be interesting to study the effects of other types of SOC, e.g., the Dresselhaus SOC, in the Kondo lattice model. 

In this work we study the noncentrosymmetric Kondo lattice model, while fully incorporating the effect of conduction electrons subject to the SOC, by virtue of a recently-developed efficient numerical simulation technique~\cite{Barros2013, Ozawa2016, Ozawa2017prl, Ozawa2017prb}. 
We introduce the SOCs of Rashba and Dresselhaus types, whose coupling constants are denoted as $\alpha$ and $\beta$, respectively. 
The Rashba SOC stems from breaking of the mirror symmetry, e.g., at the interface of a heterostructure, while the Dresselhaus SOC from breaking of the space inversion symmetry in a bulk crystal structure, e.g., in the zincblende structure.
%

Specifically we focus on three cases on a square lattice: (i) the case with both the Rashba and Dresselhaus SOCs with equal strength ($\alpha=\beta\neq0$), (ii) the case with only the Rashba SOC ($\alpha\neq0, \beta=0$), and (iii) the case with only the Dresselhaus SOC ($\alpha=0, \beta\neq0$).
The case (i) was discussed to stabilize a peculiar spin texture called the persistent spin helix~\cite{Bernevig2006}.
The situation was realized on heterostructures of zincblende-type semiconductor GaAs~\cite{Koralek2009}, and a long-living transient spin helix was observed by spin injection, e.g., through optical means~\cite{Koralek2009, Walser2012}, which may find applications to spintronics and quantum information.
In our work we treat a localized spin system coupled with conduction electrons characterized with $\alpha=\beta$, and find out sextuple-$Q$ orderings reflecting the peculiar spin-split Fermi surface. 
This situation might be potentially applied to the semiconductor quantum wells doped with magnetic impurities, though in our model the magnetic moments are positioned at every site.
The case (ii) belongs to $C_{4v}$ point group symmetry, which would be a more general and common system with broken mirror symmetry at heterointerfaces. 
In the case (ii) we discover multiple-$Q$ orderings distinct from those discussed in localized spin systems under the same symmetry.
Meanwhile, the case (iii) belongs to $D_{2d}$ point group symmetry, which is also widely encountered, not only in nonmagnetic materials like zincblende- and chalcopyrite-type semiconductors~\cite{Chen2011} but also in itinerant magnets such as a family of Heusler compounds~\cite{Nayak2017}. 
In the case (iii) we also find multiple-$Q$ orderings, which are related with those in the case (ii) by a simple global rotation.

The remaining of the paper is organized as follows.
In Sec. \ref{sec:model} we introduce the Kondo lattice model with the Rashba and Dresselhaus SOCs and derive an effective spin interactions given by the bare magnetic susceptibility.
We also discuss a unique spin-dependent gauge transformation applicable to the $\alpha=\beta$ case as well as the exchange between $\alpha$ and $\beta$. 
In Sec. \ref{sec:method} we explain the details of the numerical simulation and variational calculations.
The results are described in Sec. \ref{sec:result}. 
We devote Secs. \ref{sec:PSH}-\ref{sec:Dresselhaus} to the aforementioned three cases (i)-(iii) with different types of SOCs, respectively.
In these sections we discuss the magnetic orderings obtained by the simulation, comparing them with the bare magnetic susceptibility and the results of the variational calculations.
Finally, in Sec. \ref{sec:concl}, we summarize our results.

\section{Model}\label{sec:model}
\subsection{Hamiltonian}
In this paper we study a Kondo lattice model on a square lattice with Rashba and Dresselhaus SOCs.
The Hamiltonian is given by 
\begin{equation}
\begin{split}
\mathcal{H}&=-\sum_{jj's}t_{jj'}c_{js}^\dagger c_{j's}+\sum_{jj'ss'}i{\mathbf g}_{jj'}\cdot c_{js}^\dagger{\bm\sigma}_{ss'}c_{j's'}\\
&-J\sum_{jss'}{\mathbf S}_j\cdot c_{js}^\dagger{\bm\sigma}_{ss'}c_{js'}.
\end{split}
\label{eq:model}
\end{equation}
Here $c_{js}$ ($c_{js}^\dagger$) is the electron annihilation (creation) operator at site $j$ with spin $s(=\uparrow, \downarrow)$, and ${\mathbf S}_j={}^{\rm t}\!(S^x_j,S^y_j,S^z_j)$ describes a localized spin at site $j$, which is treated as a classical spin with the normalized length $|{\mathbf S}_j|=1$ for simplicity.
${\bm\sigma}$ is a vector of Pauli matrices, defined as ${\bm\sigma}={}^{\rm t}\!(\sigma_x,\sigma_y,\sigma_z)$.
$t_{jj'}$ represents the hopping amplitude of electrons form site $j'$ to site $j$ ($t_{jj'}=t_{j'j}$), and $J$ the spin-charge--coupling strength.
The SOCs are implemented in the second term, in which ${\mathbf g}_{jj'}$ reads 
\begin{align} 
{\mathbf g}_{jj'} = \left(
    \begin{array}{ccc}
      -\alpha_{jj'}e_{jj'}^y+\beta_{jj'}e_{jj'}^x\\
      \alpha_{jj'}e_{jj'}^x-\beta_{jj'}e_{jj'}^y\\
      0\\
    \end{array}
  \right).
\end{align}
Here $\alpha_{jj'}$ and $\beta_{jj'}$ denote the strength of Rashba and Dresselhaus SOCs, respectively, which work on an electron hopping between the sites $j$ and $j'$ ($\alpha_{jj'}=\alpha_{j'j}$ and $\beta_{jj'}=\beta_{j'j}$). 
${\mathbf e}_{jj'}=(e_{jj'}^x, e_{jj'}^y)$ is a normalized displacement vector from $j$ to $j'$, represented as ${\mathbf e}_{jj'}=({\mathbf r}_{j'}-{\mathbf r}_j)/|{\mathbf r}_{j'}-{\mathbf r}_j|$ with lattice position vectors ${\mathbf r}_j$ and ${\mathbf r}_{j'}$; we denote ${\mathbf r}_j=(n_j,m_j)$, where $n_j$ and $m_j$ are integers with the unit lattice constant.
In the following calculations, we consider the electron hopping processes between the nearest-neighbor (NN) sites and between the third-nearest-neighbor (TNN) sites.
We denote the hopping amplitudes $t_{jj'}$ between the NN and TNN sites as $t$ and $t_3$, respectively. 
Likewise, we represent the Rashba (Dresselhaus) SOC $\alpha_{jj'}$ ($\beta_{jj'}$) between the NN and TNN sites as $\alpha$ ($\beta$) and $\alpha_3$ ($\beta_3$), respectively.
In the following we take $t$ as energy unit $(t=1)$. 

In the momentum-space representation the Hamiltonian in Eq.~(\ref{eq:model}) is described as 
\begin{equation}
\mathcal{H}=\sum_{{\mathbf k}ss'}c_{{\mathbf k}s}^\dagger H^0_{ss'}({\mathbf k}) c_{{\mathbf k}s'}-J\sum_{{\mathbf k}{\mathbf q}ss'}{\mathbf S}_{\mathbf q}\cdot c_{{\mathbf k}s}^\dagger{\bm\sigma}_{ss'}c_{{\mathbf k}+{\mathbf q}s'}.
\label{eq:model_krep}
\end{equation}
Here $c_{{\mathbf k}s}$ is defined by the Fourier transform of $c_{js}$ as $c_{{\mathbf k}s} \equiv \frac{1}{\sqrt{N}} \sum_{j} e^{-i{\mathbf k}\cdot{\mathbf r}_j} c_{js}$, where $N=L^2$ is the number of sites ($L$: the linear dimension of the system).
${\mathbf S}_{\mathbf q}$ is the Fourier transform of ${\mathbf S}_j$ defined as
\begin{equation}
\mathbf{S}_\mathbf{q}=\frac{1}{N}\sum_je^{i\mathbf{q}\cdot\mathbf{r}_j}\mathbf{S}_j,
\label{eq:spinq}
\end{equation}
in which the spin normalization ($|\mathbf{S}_j|=1$) leads to the sum constraint of $\sum_{\mathbf q}\sum_\rho|S^\rho_{\mathbf q}|^2=1$ ($\rho=x, y, z$).
$H^0({\mathbf k})$ is a $2\times2$ matrix defined as
\begin{equation}
H^0({\mathbf k})=\epsilon_{\mathbf k}^0I+{\mathbf d}_{\mathbf k}\cdot{\bm\sigma},
\end{equation}
in which $I$ is the identity matrix, and $\epsilon_{\mathbf k}^0$ and ${\mathbf d}_{\mathbf k}$ are given by
\begin{equation}
\epsilon_{\mathbf k}^0=-2t(\cos k_x+\cos k_y)-2t_3(\cos 2k_x+\cos 2k_y)
\end{equation}
and
\begin{equation}
{\mathbf d}_{\mathbf k}=2\left(
    \begin{array}{ccc}
      \alpha\sin k_y-\beta\sin k_x+\alpha_3\sin 2k_y-\beta_3\sin 2k_x\\
      -\alpha\sin k_x+\beta\sin k_y-\alpha_3\sin 2k_x+\beta_3\sin 2k_y\\
      0\\
    \end{array}
  \right). 
\end{equation}

\subsection{Generalized RKKY interaction}\label{sec:rkky}
To get insight into the magnetic instability by the spin-charge coupling in the weak $J$ regime, it is useful to derive an effective spin Hamiltonian by the second-order perturbation analysis on the Hamiltonian in Eq.~(\ref{eq:model_krep}) with respect to $J$~\cite{Hayami2018}.
This gives a generalization of the Ruderman-Kittel-Kasuya-Yosida (RKKY) interactions~\cite{Ruderman1954,Kasuya1956,Yosida1957}.
The effective Hamiltonian reads
\begin{align} 
\mathcal{H}^{\mathrm{eff}}=-J^2N\sum_{\mathbf q}\sum_{\rho\rho'}S_{\mathbf q}^\rho\chi^{\rho\rho'}_{\mathbf q} (S_{\mathbf q}^{\rho'})^*,
\label{eq:rkky}
\end{align}
in which the bare magnetic susceptibility $\chi_{\mathbf q}^{\rho\rho'}$ is obtained as
\begin{equation} 
\chi_{\mathbf q}^{\rho\rho'}=\frac{T}{N}\sum_{\omega_n}\sum_{\mathbf k}{\rm tr}\left[G_0({\mathbf k}, i\omega_n)\sigma_\rho G_0({\mathbf k}+{\mathbf q}, i\omega_n)\sigma_{\rho'}\right],
\label{eq:chiqgreen}
\end{equation}
by using the noninteracting $2\times 2$ Green function $G_0({\mathbf k}, i\omega_n)=1/(i\omega_n-H^0({\mathbf k})+\mu)$; ${\omega_n}$ represents the Matsubara frequency and $\mu$ is the chemical potential.
More explicitly, Eq.~(\ref{eq:chiqgreen}) is written down as 
\begin{equation} 
\begin{split} 
\chi_{\mathbf q}^{\rho\rho'}=-\frac{1}{N}\sum_{\mathbf k}\sum_{\tau\tau^\prime}\braket{{\mathbf k}\tau|\sigma_\rho|{\mathbf k}+{\mathbf q}\tau^\prime}\braket{{\mathbf k}+{\mathbf q}\tau^\prime|\sigma_{\rho'}|{\mathbf k}\tau}\\
\times\frac{f(\epsilon_{{\mathbf k}\tau})-f(\epsilon_{{\mathbf k}+{\mathbf q}\tau^\prime})}{\epsilon_{{\mathbf k}\tau}-\epsilon_{{\mathbf k}+{\mathbf q}\tau^\prime}}.
\label{eq:chiqsimple}
\end{split} 
\end{equation}
Here $\epsilon_{{\mathbf k}\tau}$ and $\ket{{\mathbf k}\tau}$ are the eigenvalue and eigenstate of $H^0({\mathbf k})$ with the band index $\tau$.
$f(\epsilon)$ is the Fermi distribution function expressed as $f(\epsilon)=1/(1+e^{(\epsilon-\mu)/k_{\rm B}T})$, where $k_{\mathrm{B}}$ is the Boltzmann constant and $T$ is the temperature. 

In the presence of SOC, in general, the bare magnetic susceptibility $\chi_{\mathbf q}^{\rho\rho'}$ in Eq.~(\ref{eq:chiqsimple}) has nonzero off-diagonal components.
To examine the dominant magnetic instability, therefore, it is useful to diagonalize the effective spin Hamiltonian in Eq.~(\ref{eq:rkky}) in the form: 
\begin{align} 
\mathcal{H}^{\mathrm{eff}}=-J^2N\sum_{\mathbf q}\sum_{\xi}\lambda^\xi_{\mathbf q} |S'^\xi_{\mathbf q}|^2.
\label{eq:rkky_diag}
\end{align}
Here we define the eigenvalues and eigenvectors of $\chi^{\rho\rho'}_{\mathbf q}$ in Eq.~(\ref{eq:chiqsimple}) as $\lambda^\xi_{\mathbf q}$ and ${\mathbf u}^\xi_{\mathbf q}$ ($\xi=1-3$), respectively, formulated as
\begin{equation} 
\chi_{\mathbf q}{\mathbf u}^\xi_{\mathbf q}=\lambda^\xi_{\mathbf q}{\mathbf u}^\xi_{\mathbf q}.
\label{eq:chiq_eigen} 
\end{equation} 
Note that we sort the eigenvalues as $\lambda^1_{\mathbf q}\geqq\lambda^2_{\mathbf q}\geqq\lambda^3_{\mathbf q}$.
${\mathbf S}'_{\mathbf q}$ is a transformed spin Fourier component, given by ${\mathbf S}_{\mathbf q}=U_{\mathbf q}^*{\mathbf S}'_{\mathbf q}$ with $U_{\mathbf q}=[{\mathbf u}^1_{\mathbf q}, {\mathbf u}^2_{\mathbf q}, {\mathbf u}^3_{\mathbf q}]$.
Note the sum constraint also holds for ${\mathbf S}'_{\mathbf q}$ as $\sum_{\mathbf q}\sum_{\xi}|S'^\xi_{\mathbf q}|^2=1$.

The diagonalized form of the effective spin Hamiltonian in Eq.~(\ref{eq:rkky_diag}) gives us important information on magnetic instability. 
Suppose the largest eigenvalue $\lambda^1_{\mathbf q}$ takes the maxima at a set of wave vectors $\left\{{\mathbf Q}_\nu\right\}$. 
Then, under the sum constraint of $\sum_{\mathbf q}\sum_{\xi}|S'^\xi_{\mathbf q}|^2=1$, we find that the largest energy gain of the RKKY Hamiltonian in Eq.~(\ref{eq:rkky_diag}) is earned for multiple- or single-$Q$ magnetic orderings characterized with the wave vectors $\left\{{\mathbf Q}_\nu\right\}$ with the corresponding spin Fourier components of ${\mathbf S}_{{\mathbf Q}_\nu}\propto ({\mathbf u}^1_{{\mathbf Q}_\nu})^*$ (see also Sec.~\ref{sec:KPMLD}).
Therefore, analyzing the ${\mathbf q}$ profile of $\lambda^1_{\mathbf q}$ is important to figure out the inherent magnetic instability in the weak $J$ regime.

Meanwhile, it should be noted that the generalized RKKY interactions leave degeneracy; the single- and multiple-$Q$ orderings specified by $\left\{{\mathbf Q}_\nu\right\}$ and the corresponding modes ${\mathbf S}_{{\mathbf Q}_\nu}\propto ({\mathbf u}^1_{{\mathbf Q}_\nu})^*$ are energetically degenerate.
Higher-order contributions play a crucial role in selecting out the lowest-energy magnetic state, as demonstrated in the absence of SOC~\cite{Akagi2012,Hayami2014,Ozawa2016}.
This motivates us to study the original model in Eq.~(\ref{eq:model}) or (\ref{eq:model_krep}) by numerical simulation that treats the spin-charge coupling and the SOC on an equal footing.

\subsection{Spin-dependent gauge transformation for $\alpha=\beta$}\label{sec:gauge}
In the case of $\alpha=\beta$ with only the NN terms ($t_3=\alpha_3=\beta_3=0$), the Fermi surfaces have peculiar properties~\cite{Bernevig2006}.
The Fermi surfaces, which have spin degeneracy in the absence of SOC, are unidirectionally split along the [$\bar{1}10$] direction by the SOC, and moreover, all the states in each Fermi surface have the same spin polarization parallel or antiparallel to the [$110$] direction [for example, see Fig.~\ref{fig:PSH_Fermisurface}(c)].
The shift vector connecting the spin-split Fermi surfaces, ${\mathbf Q}^{\rm s}$, is given by ${\mathbf Q}^{\rm s}=2{\rm tan}^{-1}(\sqrt{2}\alpha)(-1, 1)$.
Importantly, this peculiar nature of the Fermi surfaces indicates that the SOC with $\alpha=\beta$ can be effectively taken away through a certain spin-dependent gauge transformation, which adds or subtracts half of the shift vector, ${\mathbf Q}^{\rm s}/2$, to or from the electron momenta, depending on the spin directions~\cite{Bernevig2006}.
For the annihilation operators, the gauge transformation can be formulated as 
\begin{equation} 
\tilde{{\mathbf c}}_j=\left(
    \begin{array}{cc}
      e^{i{\mathbf Q}^{\rm s}\cdot{\mathbf r}_j/2}&0\\
      0&e^{-i{\mathbf Q}^{\rm s}\cdot{\mathbf r}_j/2}\\
    \end{array}
  \right)V_0{\mathbf c}_j,
\end{equation} 
where ${\bf c}_j = {}^{\rm t}(c_{j\uparrow} ,c_{j\downarrow})$ and 
\begin{equation} 
V_0 = \frac{1}{\sqrt{2}}\left(
    \begin{array}{cc}
      e^{i\frac{\pi}{4}}&1\\
      e^{i\frac{\pi}{4}}&-1\\
    \end{array}
  \right). 
\end{equation} 
This transformation adds spin-dependent gauges with the quantization axis to [$110$]. 
Then, by using the newly-defined annihilation and creation operators, $\tilde{{\mathbf c}}_j$ and $\tilde{{\mathbf c}}^\dagger_j$, the original Hamiltonian in Eq.~(\ref{eq:model}) for $\alpha=\beta$ with only the NN terms ($t=1$) is written into the form with effectively-vanishing SOC:
\begin{equation} 
\mathcal{H}=-\sqrt{1+2\alpha^2}\sum_{jj's}{\tilde c}_{js}^\dagger {\tilde c}_{j's}-J\sum_{jss'}{\tilde {\mathbf S}}_j\cdot {\tilde c}_{js}^\dagger{\bm\sigma}_{ss'}{\tilde c}_{js'}.
\label{eq:PSH_trasformedmodel}
\end{equation}
Here the new spin frame ${\tilde {\mathbf S}}_j$ is defined through the rotation by the amount of ${\mathbf Q}^{\rm s}\cdot{\mathbf r}_j$ along the [110] direction on the original spin frame as 
\begin{equation} 
{\tilde{\mathbf S}}_j =
\left(
    \begin{array}{ccc}
      \cos{\mathbf Q}^{\rm s}\cdot{\mathbf r}_j&\sin{\mathbf Q}^{\rm s}\cdot{\mathbf r}_j&0\\
      -\sin{\mathbf Q}^{\rm s}\cdot{\mathbf r}_j&\cos{\mathbf Q}^{\rm s}\cdot{\mathbf r}_j&0\\
     0&0&1\\
    \end{array}
  \right)
\left(
    \begin{array}{ccc}
      0&0&1\\
      \frac{1}{\sqrt{2}}&-\frac{1}{\sqrt{2}}&0\\
      \frac{1}{\sqrt{2}}&\frac{1}{\sqrt{2}}&0\\
    \end{array}
  \right)
{\mathbf S}_j.
\label{eq:equalRD_spintransform}
\end{equation}
These analyses imply that the magnetic orderings for $\alpha=\beta\neq0$ are related to those without SOC through the site-dependent rotation in Eq.~(\ref{eq:equalRD_spintransform}).
We use this property in the discussion in Sec.~\ref{sec:PSH}.

\subsection{Exchange between $\alpha$ and $\beta$}\label{sec:exchange}
We also remark that the exchange between $\alpha$ and $\beta$ leads to a simple uniform rotation of magnetic orderings.
By applying a $\pi$ rotation along the [$110$] axis in spin space for conduction electrons as given by 
\begin{equation}
\overline{\mathbf c}_j= \exp\left(-i\frac{\pi}{2}\frac{\sigma_x+\sigma_y}{\sqrt{2}}\right){\mathbf c}_j, 
\label{eq:exchangeRD_spintransform_c}
\end{equation}
and likewise to the spin frame for the localized spins as
\begin{equation}
\overline{\mathbf S}_j = \left(
    \begin{array}{ccc}
      0&1&0\\
      1&0&0\\
     0&0&-1\\
    \end{array}
  \right){\mathbf S}_j,
\label{eq:exchangeRD_spintransform}
\end{equation}
the original Hamiltonian in Eq.~(\ref{eq:model}) is transformed to the one with exchanged $\alpha_{jj'}$ and $\beta_{jj'}$: 
\begin{equation} 
\begin{split} 
\mathcal{H}&=-\sum_{jj's}t_{jj'}\overline{c}_{js}^\dagger\overline{c}_{j's}\\
&+\sum_{jj'ss'}i{\mathbf g}_{jj'}(\alpha_{jj'}\leftrightarrow\beta_{jj'})\cdot\overline{c}_{js}^\dagger{\bm\sigma}_{ss'}\overline{c}_{j's'}\\
&-J\sum_{jss'}\overline{\mathbf S}_j\cdot\overline{c}_{js}^\dagger{\bm\sigma}_{ss'}\overline{c}_{js'}.
\end{split} 
\end{equation}
This indicates that the magnetic orderings for the Rashba-only case in Sec.~\ref{sec:Rashba} also applies to the Dresselhaus-only case in Sec.~\ref{sec:Dresselhaus} through the global rotations in Eqs.~(\ref{eq:exchangeRD_spintransform_c}) and (\ref{eq:exchangeRD_spintransform}).
We utilize this nature in Sec.~\ref{sec:Dresselhaus}.

\section{Method}\label{sec:method}
\subsection{KPM-LD}\label{sec:KPMLD}
To reveal the ground-state magnetic orderings for the Kondo lattice model with Rashba and Dresselhaus SOCs we employ a state-of-the-art large-scale numerical simulation combining the kernel polynomial method (KPM)~\cite{Weisse2006} with Langevin dynamics (LD)~\cite{Barros2013}.
This recently-developed method, called KPM-LD, costs only $O(N)$ ($N$: number of lattice sites), allowing us to run the simulation for the system sizes of up to $\sim 10^4$ sites. 
Here we employ the modified version of the KPM-LD~\cite{Ozawa2017prb} making use of a probing method~\cite{Tang2012} and the stochastic Landau-Lifshitz-Gilbert equation in the LD.

We perform the KPM-LD at zero temperature on the square lattice of $N=96^2$.
In the KPM, we expand the density of states in a series of Chebyschev polynomials up to the 2000th order, in which 144 random vectors are chosen by a probing technique~\cite{Tang2012} for calculation of the Chebyschev moments.

In Sec.~\ref{sec:PSH} the KPM-LD is initiated from a random spin configuration, aiming at an unbiased search for the ground state.
On the other hand, in Sec.~\ref{sec:Rashba}, we start the KPM-LD from some given ansatzes for the configuration of localized spins, because for $\alpha\neq0$ and $\beta=0$ we found that random configurations often fail to converge to a homogeneous state and end up with a mixing of different ordering domains. This can be attributed to keen energy competitions of multiple magnetic orders originating in a considerable number of sharp peaks in $\lambda^1_{\mathbf q}$ [see Fig.~\ref{fig:Rashba_Fermisurface}(f)].
Consequently, in Sec.~\ref{sec:Rashba} we use the KPM-LD as an ``ansatz optimizer" rather than an unbiased simulation. 

Below we describe how we prepare the initial ansatzes used in Sec.~\ref{sec:Rashba}.
The ansatzes we employ are single-$Q$ helical states that maximize the energy gain of the generalized RKKY Hamiltonian in Eq.~(\ref{eq:rkky_diag}) and multiple-$Q$ superpositions of them.
As discussed later in Sec.~\ref{sec:Rashba}, for $\alpha\neq0$ and $\beta=0$, $\lambda^1_{\mathbf q}$ takes the largest value at four wave vectors denoted as $\mathbf{Q}^a_\nu$ $(\nu=1-4)$ among all the characteristic wave vectors [see Fig.~\ref{fig:Rashba_Fermisurface}(d)].
$\left\{\mathbf{Q}^a_\nu\right\}$ are related with each other by $C_4$ and $\sigma_v$ symmetry operations. 
Moreover, the $C_{4v}$ symmetry dictates that the corresponding eigenvectors ${\mathbf u}^1_{{\mathbf Q}^a_\nu}$ are simply described as 
\begin{subequations}
\begin{eqnarray} 
{\mathbf u}^1_{\mathbf{Q}^a_1}&=&{}^{\rm t}\!(u_{x},u_{y},iu_{z}),\\
{\mathbf u}^1_{\mathbf{Q}^a_2}&=&{}^{\rm t}\!(u_{y},u_{x},iu_{z}),\\
{\mathbf u}^1_{\mathbf{Q}^a_3}&=&{}^{\rm t}\!(-u_{y},u_{x},iu_{z}),\\
{\mathbf u}^1_{\mathbf{Q}^a_4}&=&{}^{\rm t}\!(-u_{x},u_{y},iu_{z}),
\end{eqnarray}
\end{subequations}
where $u_x$, $u_y$, and $u_z$ are real numbers.
As mentioned in Sec.~\ref{sec:rkky}, multiple-$Q$ orderings maximizing the RKKY energy gain in Eq.~(\ref{eq:rkky_diag}) under the sum constraint of $\sum_{\mathbf q}\sum_\rho|S^\rho_{\mathbf q}|^2=1$ are characterized with the spin Fourier components of ${\mathbf S}_{{\mathbf Q}^a_\nu}\propto ({\mathbf u}^1_{{\mathbf Q}^a_\nu})^*$.
As a result, in this Rashba-only case the multiple-$Q$ states are given by superpositions of symmetry-related helices with the spin rotation plane perpendicular to the $xy$-plane, which are given by 
\begin{equation} 
{\mathbf S}_j = \hat{N}\sum_{\nu=1}^4A_\nu\left(
    \begin{array}{ccc}
      u^{1x}_{{\mathbf Q}^a_\nu}\cos \mathbf{Q}^a_{\nu}\cdot\mathbf{r}_j\\
      u^{1y}_{{\mathbf Q}^a_\nu}\cos \mathbf{Q}^a_{\nu}\cdot\mathbf{r}_j\\
      -u^{1z}_{{\mathbf Q}^a_\nu}\sin \mathbf{Q}^a_{\nu}\cdot\mathbf{r}_j\\
    \end{array}
  \right).
\label{eq:rashba_ansatz}
\end{equation}
Here the sum constraint $\sum_\nu A_\nu^2=4$ holds for $A_\nu$ and $\hat{N}$ represents the normalization factor so that $|{\mathbf S}_j|=1$.
We note that without the normalization factor $\hat{N}$ all the ansatzes described by Eq.~(\ref{eq:rashba_ansatz}) gain the same amount of the RKKY energy in Eq.~(\ref{eq:rkky_diag}).
Among those multiple-$Q$ orderings we take double- or single-$Q$ orderings for the initial ansatzes in the KPM-LD for simplicity. 
For the double-$Q$ orderings, we set $A_{\nu}=\sqrt{2}$ for $\nu=1$ and $2$, or $\nu=1$ and $3$.
Likewise, for the single-$Q$ ordering, we set $A_{\nu}=2$ for one $\nu$ and otherwise $A_{\nu}=0$.

In Sec.~\ref{sec:PSH} we employ the periodic boundary condition as in the previous works~\cite{Ozawa2016, Ozawa2017prl}, while in Sec.~\ref{sec:Rashba} we use the open boundary condition. 
This is because in the latter case it turns out that the KPM-LD yields incommensurate magnetic orderings with a large magnetic unit cell [see Fig.~\ref{fig:Rashba_spin}(c)], which would be attributed to the wave vectors with the largest $\lambda^1_{\mathbf q}$, $\left\{{\mathbf Q}^a_\nu\right\}$, deviating from commensurate wave vectors [see Fig.~\ref{fig:Rashba_Fermisurface}(d)].
In order to exclude the boundary effects we extract the square with $64^2$ sites in the middle of the whole system with $96^2$ sites for analyzing the spin textures.

For the spin textures obtained in the KPM-LD we calculate $|{\mathbf S}_{\mathbf q}|$ [see Eq.~(\ref{eq:spinq})], which is proportional to the square root of the spin structure factor. 
We also compute the spin scalar chirality $\kappa_p$ for each square plaquette $p$ as 
\begin{equation} 
\kappa_p=\frac{1}{4}(\mathbf{S}_j\cdot\mathbf{S}_k\times\mathbf{S}_l+\mathbf{S}_k\cdot\mathbf{S}_l\times\mathbf{S}_m+\mathbf{S}_l\cdot\mathbf{S}_m\times\mathbf{S}_j+\mathbf{S}_m\cdot\mathbf{S}_j\times\mathbf{S}_k),
\end{equation}
where the sites $j$, $k$, $l$, and $m$ correspond to the bottom-left, bottom-right, top-right and top-left vertices of the square plaquette $p$, respectively.
In the same manner as $|{\mathbf S}_{\mathbf q}|$, we define $|\kappa_{\mathbf q}|=|\frac{1}{N}\sum_p e^{i{\mathbf q}\cdot{\mathbf r}_p}\kappa_p|$.
 
\subsection{Variational calculation}\label{sec:variational}
In Secs.~\ref{sec:PSH} and \ref{sec:Rashba} we also perform variational calculations.
For given spin configurations we calculate the total energy by using the exact diagonalization of the one-body Hamiltonian and compare the values to determine the ground state.
The calculations are done for the system sizes of $N=96^2$ and $480^2$.

\section{Results}\label{sec:result}
\subsection{Case with $\alpha=\beta$}\label{sec:PSH}
\begin{figure}[!htb]
\begin{center}
\includegraphics[width=\linewidth,clip]{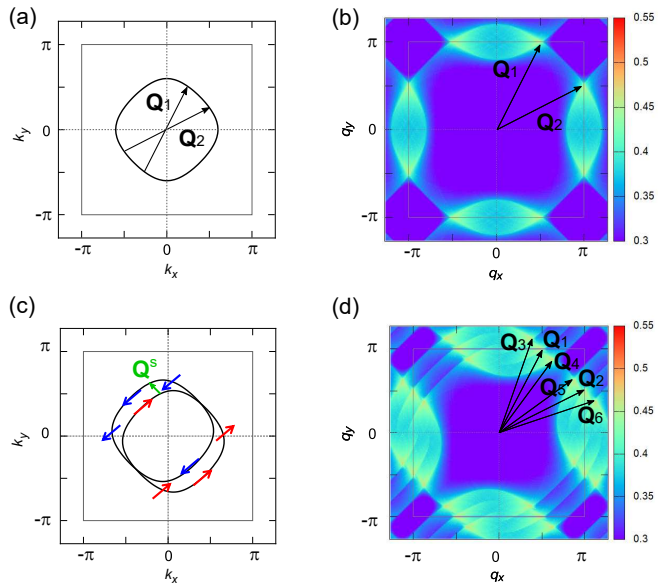}
\end{center}
\caption{Fermi surfaces and bare magnetic susceptibilities for $\mu\sim-1.4$ (near quarter filling) (a,b) without the SOC ($\alpha=\beta=0$) and (c,d) with the equally-strong Rashba and Dresselhaus SOCs ($\alpha=\beta=0.2$).
(a,c) and (b,d) show the Fermi surfaces and the largest eigenvalues of the bare magnetic susceptibility $\lambda^1_{\mathbf q}$ [see Eq.~(\ref{eq:chiq_eigen})], respectively. Note that all the TNN terms are set to zero ($t_3=\alpha_3=\beta_3=0$).}
\label{fig:PSH_Fermisurface}
\end{figure}

\begin{figure*}[!htb]
\includegraphics[width=\linewidth,clip]{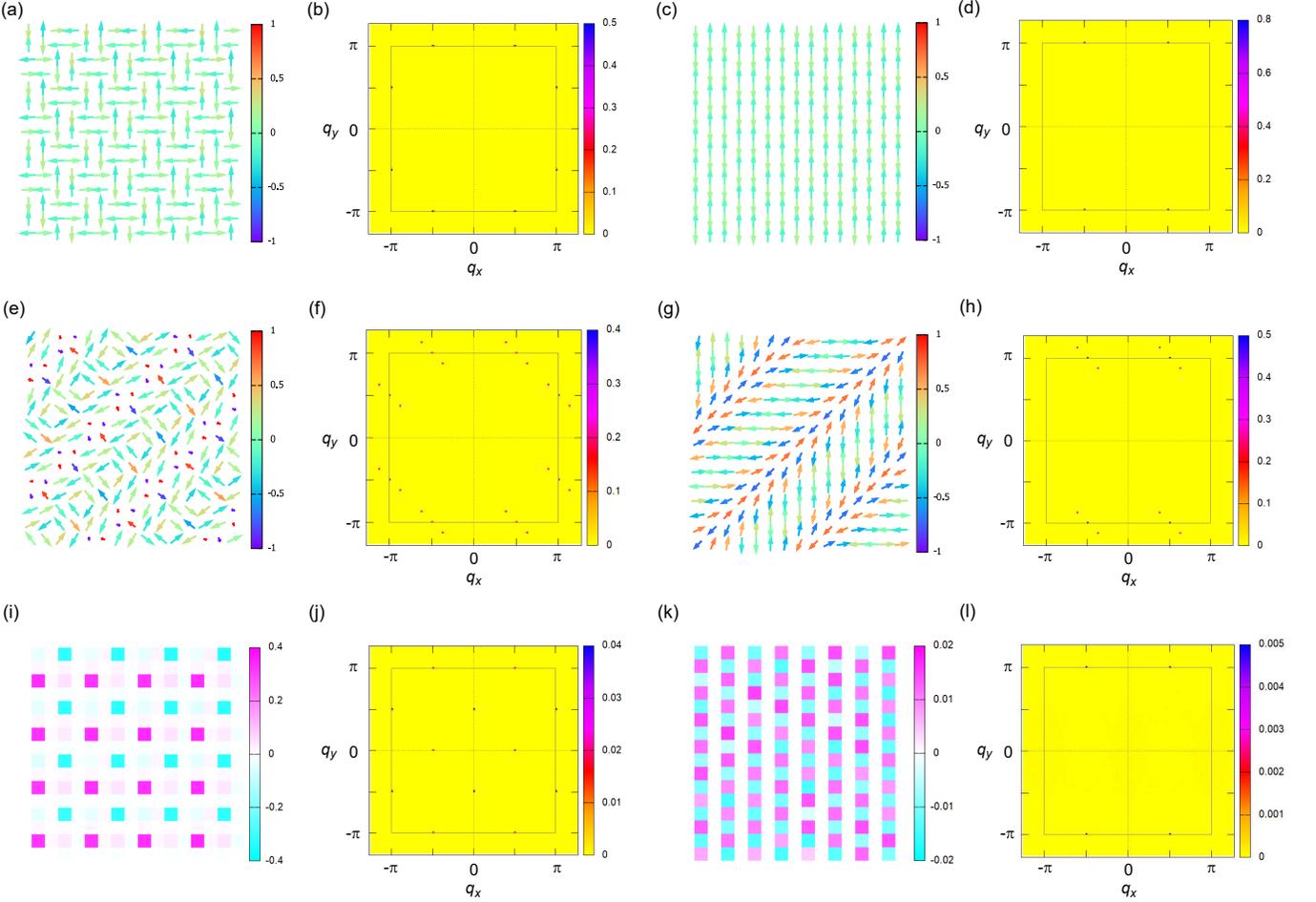}
\caption{The results of the KPM-LD simulations for $\alpha=\beta=0$ and $\alpha=\beta=0.2$.
(a-d) correspond to $\alpha=\beta=0$.
(a) and (b) represent a real-space spin texture and the norm of its Fourier transform, $|{\mathbf S}_{\mathbf q}|$, of $2Q$-uudd for $J=0.1$, respectively, while (c) and (d) correspond to those of $1Q$-uudd for $J=0.4$. 
(e-l) correspond to $\alpha=\beta=0.2$.
(e) and (f) are a spin texture and $|{\mathbf S}_{\mathbf q}|$ of $6Q$ for $J=0.1$, while (g) and (h) display those of $3Q$ for $J=0.4$.
(i) shows the real-space pattern of the spin scalar chirality for $6Q$ corresponding to (e), with the absolute value of its Fourier transform, $|\kappa_{\mathbf q}|$, represented in (j).
(k) and (l) are those for $3Q$. 
The real-space textures of spin and scalar chirality in (a), (c), (e), (g), (i), and (k) are shown for a part of the whole system with $N=96^2$ system for clarity.
In (a), (c), (e), and (g), the arrows denote the directions of the localized spins in the $xy$ plane and the color represents the $z$ component.
}
\label{fig:PSH_spin}
\end{figure*}

First, we discuss the Fermi surface instabilities for the case with equally-strong Rashba and Dresselhaus SOCs, namely, $\alpha=\beta$, along with the case without the SOCs.
In this section we consider only the NN terms in the Hamiltonian in Eq.~(\ref{eq:model}) and neglect the TNN terms ($t_3=\alpha_3=\beta_3=0$).
Here we set the chemical potential as $\mu\sim-1.4$, corresponding to near quarter filling ($n\sim0.5$).
Figures~\ref{fig:PSH_Fermisurface}(a) and \ref{fig:PSH_Fermisurface}(b) display the Fermi surface and the largest eigenvalue of the bare magnetic susceptibility in the absence of SOC.
The Fermi surface is partially nested by the commensurate wave vectors ${\mathbf Q}_1=(\pi/2,\pi)$ and ${\mathbf Q}_2=(\pi,\pi/2)$ at this filling, as illustrated in Fig.~\ref{fig:PSH_Fermisurface}(a).
Reflecting the partial nesting, the susceptibility takes the largest value at two inequivalent positions on the edge of the Brillouin Zone (BZ), ${\mathbf Q}_1$ and ${\mathbf Q}_2$, as shown in Fig.~\ref{fig:PSH_Fermisurface}(b).

When $\alpha$ and $\beta$ are introduced with equal strength, the spin-degenerate Fermi surfaces are split along the [$\bar{1}$10] direction, each of which has the uniform spin polarization parallel or antiparallel to the [110] direction.
Figure~\ref{fig:PSH_Fermisurface}(c) shows the spin-split Fermi surfaces for $\alpha=\beta=0.2$. 
The partial nesting of the shifted Fermi surfaces yield additional maxima in $\lambda^1_{\mathbf q}$ at four wave vectors, ${\mathbf Q}_3={\mathbf Q}_1+{\mathbf Q}^{\rm s}$, ${\mathbf Q}_4={\mathbf Q}_1-{\mathbf Q}^{\rm s}$, ${\mathbf Q}_5={\mathbf Q}_2+{\mathbf Q}^{\rm s}$, and ${\mathbf Q}_6={\mathbf Q}_2-{\mathbf Q}^{\rm s}$, where ${\mathbf Q}^{\rm s}$ is the shift vector of the Fermi surfaces shown in Fig.~\ref{fig:PSH_Fermisurface}(c).
As a result, the bare magnetic susceptibility shows the largest value at totally six independent wave vectors as shown in Fig.~\ref{fig:PSH_Fermisurface}(d).

With the Fermi surface instabilities at these wave numbers in mind we discuss the spin textures obtained by the KPM-LD simulations.
We begin with the results for $J=0.1$.
In the absence of SOC we obtain the noncollinear but coplanar double-$Q$ ordering [Figs.~\ref{fig:PSH_spin}(a) and \ref{fig:PSH_spin}(b)], as reported in the previous work~\cite{Hayami2016}. 
The two wave vectors characterizing the magnetic texture are identified as ${\mathbf Q}_1$ and ${\mathbf Q}_2$, which coincide with those in Figs.~\ref{fig:PSH_Fermisurface}(a) and \ref{fig:PSH_Fermisurface}(b).
Since the spin components are modulated in the up-up-down-down manner, hereafter we refer to this double-$Q$ order as $2Q$-uudd~\cite{Hayami2016}.

On the other hand, when the Rashba and Dresselhaus SOCs are introduced with the equal strength of $\alpha=\beta=0.2$, we find a complex noncoplanar spin texture characterized with six wave vectors, as shown in Figs.~\ref{fig:PSH_spin}(e) and \ref{fig:PSH_spin}(f). 
These wave vectors coincide with the ones giving the largest value in $\lambda^1_{\mathbf q}$, $\left\{{\mathbf Q}_\nu\right\}$ ($\nu=1-6$) in Fig.~\ref{fig:PSH_Fermisurface}(d).
It is noteworthy that to the best of our knowledge there is no other theoretical or experimental report showing stabilization of any magnetic ordering with more than three wave vectors in two-dimensional systems. 
Remarkably, we find that this sextuple-$Q$ ordering ($6Q$) exhibits a checkerboard-like pattern of the spin scalar chirality [Fig.~\ref{fig:PSH_spin}(i)], characterized with multiple wave vectors specified by $(\pi/2,0)$, $(0,\pi/2)$, $(\pi,\pi/2)$, and $(\pi/2,\pi)$ [Fig.~\ref{fig:PSH_spin}(j)].

While increasing $J$ to $J=0.2$ and $0.3$, we find that the same ordering patterns are obtained in the KPM-LD: $2Q$-uudd without SOC and $6Q$ with $\alpha=\beta=0.2$. 
The result indicates that the Fermi surface instabilities govern the magnetic textures in the weak coupling regime.

For $J=0.4$, however, we find that the spin texture changes into a less complex one.
Without the SOC appears a simple single-$Q$ state composed of ${\mathbf Q}_1$ (or ${\mathbf Q}_2$, depending on the initial configuration), which is a collinear up-up-down-down ordering [Figs.~\ref{fig:PSH_spin}(c) and \ref{fig:PSH_spin}(d)].
In the same way as $2Q$-uudd, we denote this single-$Q$ order as $1Q$-uudd~\cite{Hayami2016}.
With $\alpha=\beta=0.2$ we obtain the triple-$Q$ ordering characterized by the three ordering vectors ${\mathbf Q}_1$, ${\mathbf Q}_2$, and ${\mathbf Q}_3$ (or ${\mathbf Q}_4$, ${\mathbf Q}_5$, and ${\mathbf Q}_6$) [Figs.~\ref{fig:PSH_spin}(g) and \ref{fig:PSH_spin}(h)].
The $3Q$ state also shows the density wave of the spin scalar chirality as shown in Fig.~\ref{fig:PSH_spin}(k), although it is only characterized by a single wave vector as shown in Fig.~\ref{fig:PSH_spin}(l) in contrast to four in Fig.~\ref{fig:PSH_spin}(j).

As we mentioned in Sec.~\ref{sec:gauge}, the spin-dependent gauge transformation guarantees the exact mapping of the model for $\alpha=\beta\neq0$ to the SOC-free one in Eq.~(\ref{eq:PSH_trasformedmodel}).
Hence, the magnetic orderings stabilized for $\alpha=\beta\neq0$ are related with those for $\alpha=\beta=0$ through the transformation in Eq.~(\ref{eq:equalRD_spintransform}). 
Indeed, we have confirmed that $6Q$ and $3Q$ uncovered in the KPM-LD simulations are obtained by applying the site-dependent rotation in Eq.~(\ref{eq:equalRD_spintransform}) to $2Q$-uudd and $1Q$-uudd, respectively, after certain global rotations. 

\begin{figure}[!htb]
\centering
\includegraphics[width=0.8\columnwidth]{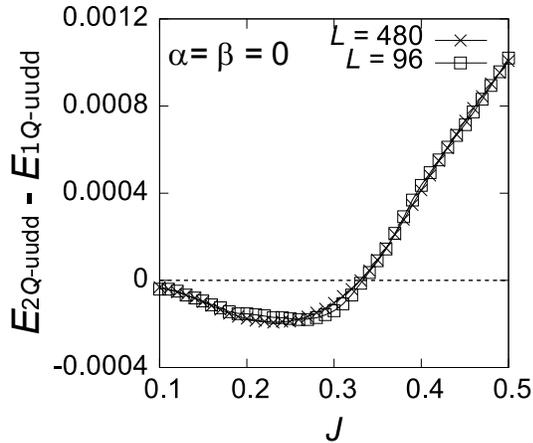}
\caption{
Energy difference between $2Q$-uudd and $1Q$-uudd at $\mu\sim -1.4$ in the absence of SOC, estimated by variational calculations for $N=L^2=96^2$ and $480^2$. 
}
\label{fig:PSH_variational}
\end{figure}

We also verified the results of the KPM-LD by variational calculations.
Figure~\ref{fig:PSH_variational} shows the energy difference between $2Q$-uudd and $1Q$-uudd obtained in the absence of SOC ($\alpha=\beta=0$).
Here we take the variational states as
\begin{equation} 
{\mathbf S}_j = \left(
    \begin{array}{ccc}
      \cos({\mathbf Q}_1\cdot{\mathbf r}_j-\frac{\pi}{4})\\
      \cos({\mathbf Q}_2\cdot{\mathbf r}_j-\frac{\pi}{4})\\
      0\\
    \end{array}
  \right), 
\end{equation}
for $2Q$-uudd, and
\begin{equation} 
{\mathbf S}_j = \left(
    \begin{array}{ccc}
     \sqrt{2}\cos({\mathbf Q}_1\cdot{\mathbf r}_j-\frac{\pi}{4})\\
      0\\
      0\\
    \end{array}
  \right), 
\end{equation} 
for $1Q$-uudd~\cite{Hayami2016}.
Note that we do not need the normalization factor for the spin lengths as the wave numbers are commensurate.
As shown in Fig.~\ref{fig:PSH_variational}, $2Q$-uudd is more stable compared to $1Q$-uudd for $J<J_{\rm c}^0\sim0.33$ and vice versa for $J>J_{\rm c}^0$. 
The variational result looks consistent with the KPM-LD results.
By using the spin-dependent gauge transformation, we can derive the critical value of $J$ for nonzero $\alpha=\beta$ as $J_{\rm c}=J^0_{\rm c}\sqrt{1+2\alpha^2}$.

\begin{figure}[!htb]
\centering
\includegraphics[width=0.9\columnwidth]{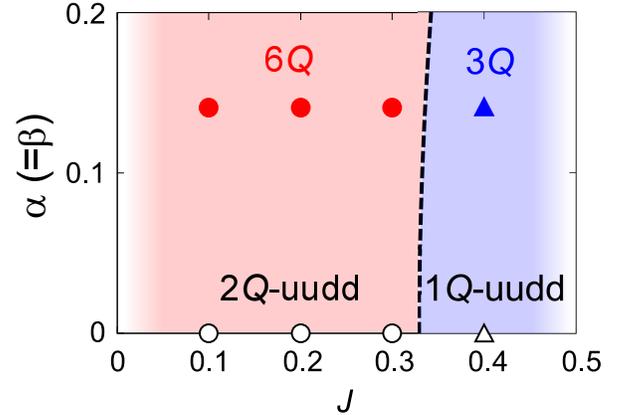}
\caption{Ground-state phase diagram for the Kondo lattice model with equally-large Rashba and Dresselhaus SOCs ($\alpha=\beta$) for $\mu\sim -1.4$, determined by the KPM-LD and variational calculations.
The circles and triangles represent the parameters for which the KPM-LD simulations are performed.
In the absence of SOC ($\alpha=\beta=0$) $2Q$-uudd is favored for $J<J_{\rm c}^0\sim0.33$, while $1Q$-uudd is stabilized for $J>J_{\rm c}^0$ (see Fig.~\ref{fig:PSH_variational}). 
For finite SOCs ($\alpha=\beta\neq0$) $6Q$ appears on the red-shaded region, while $3Q$ shows up on the blue-shaded region.
The dashed line is the phase boundary given by $J_{\rm c}=J^0_{\rm c}\sqrt{1+2\alpha^2}$.
}
\label{fig:PSH_phase_diagram}
\end{figure}

Combining the results by the KPM-LD and variational calculations, we summarize the $J$-$\alpha$ phase diagram for equally-large $\alpha$ and $\beta$ in Fig.~\ref{fig:PSH_phase_diagram}.
The red- and blue-shaded regions correspond to $6Q$ and $3Q$, respectively, and the dashed line shows the phase boundary $J_{\rm c}$ determined by the variational calculations.
The phase diagram in Fig.~\ref{fig:PSH_phase_diagram} indicates that the exotic sextuple-$Q$ orderings are stabilized in a wide parameter range of $\alpha$ and $J$ in the present spin-charge and spin-orbit coupled system. 

\subsection{Case with $\alpha\neq0$ and $\beta=0$}\label{sec:Rashba}
\begin{figure}[!htb]
\includegraphics[width=\linewidth,clip]{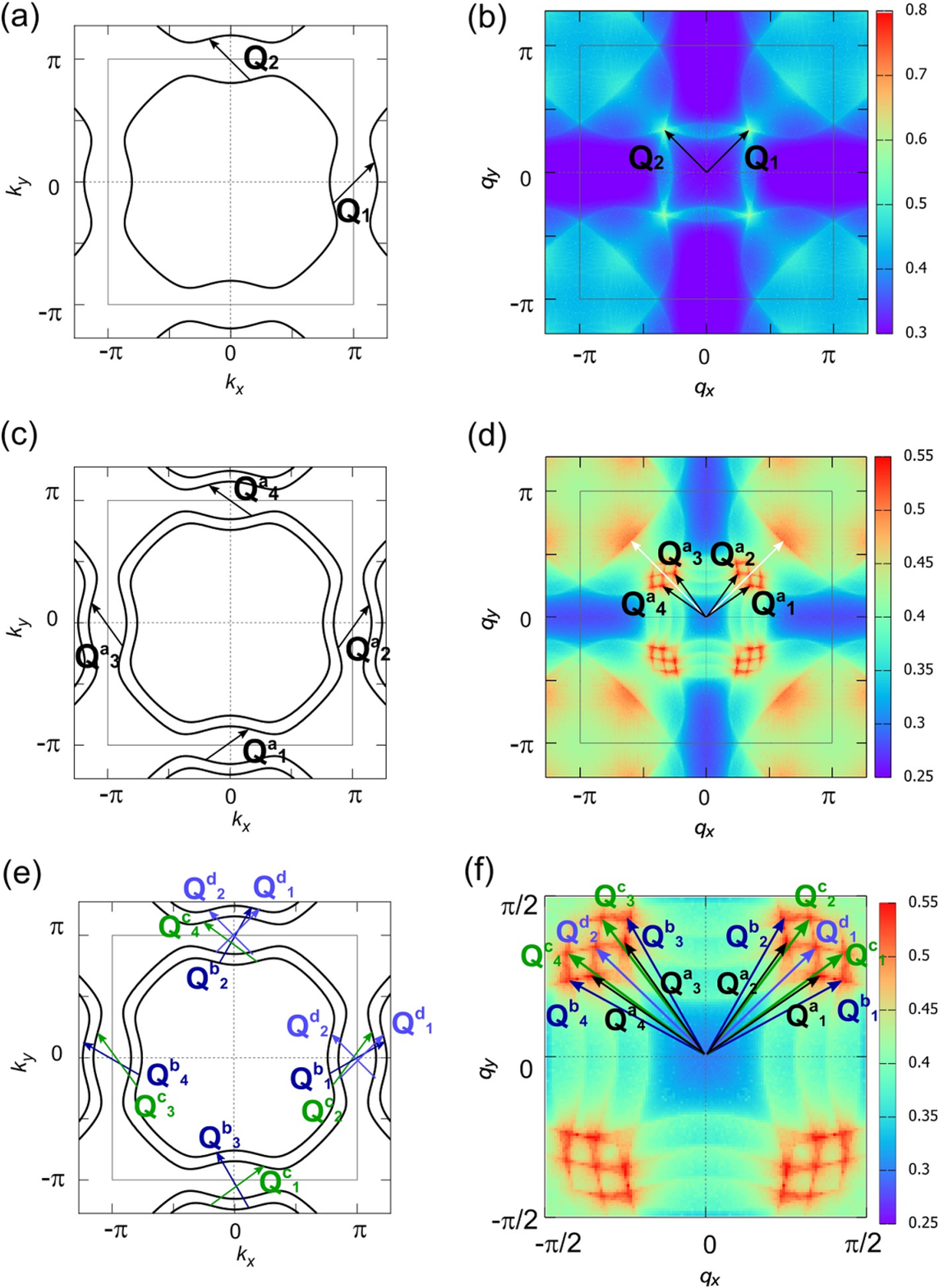}
\caption{Fermi surfaces and bare magnetic susceptibilities for $\mu=0.98$ (a,b) without the SOCs ($\alpha=\beta=0$) and (c-f) with the Rashba SOC ($\alpha=0.2, \beta=0$).
Note that the TNN terms are introduced with $t_3=-0.5$ and $\alpha_3=-0.5\alpha$.
(a,c,e) and (b,d,f) show the Fermi surfaces and the largest eigenvalues of the bare magnetic susceptibility, ${\lambda}^1_{\mathbf q}$ [see Eq. (\ref{eq:chiq_eigen})], respectively.
(f) is the magnified view of (d).
In (a) and (b) the arrows indicate the wave vectors that give the largest magnetic susceptibility in the absence of SOC.
In (c) and (d) the black arrows denote the wave vectors that give the largest ${\lambda}^1_{\mathbf q}$ in the presence of the Rashba SOC.
In (d) the white arrows correspond to the ordering vectors of $1Q'$, which is found for $J=0.2-0.4$ in the KPM-LD simulations (see Fig. \ref{fig:Rashba_phase_diagram}).
In (e) and (f) the other characteristic wave vectors, which give the comparably large ${\lambda}^1_{\mathbf q}$, are shown.
}
\label{fig:Rashba_Fermisurface}
\end{figure}

\begin{figure*}[!htb]
\includegraphics[width=\linewidth,clip]{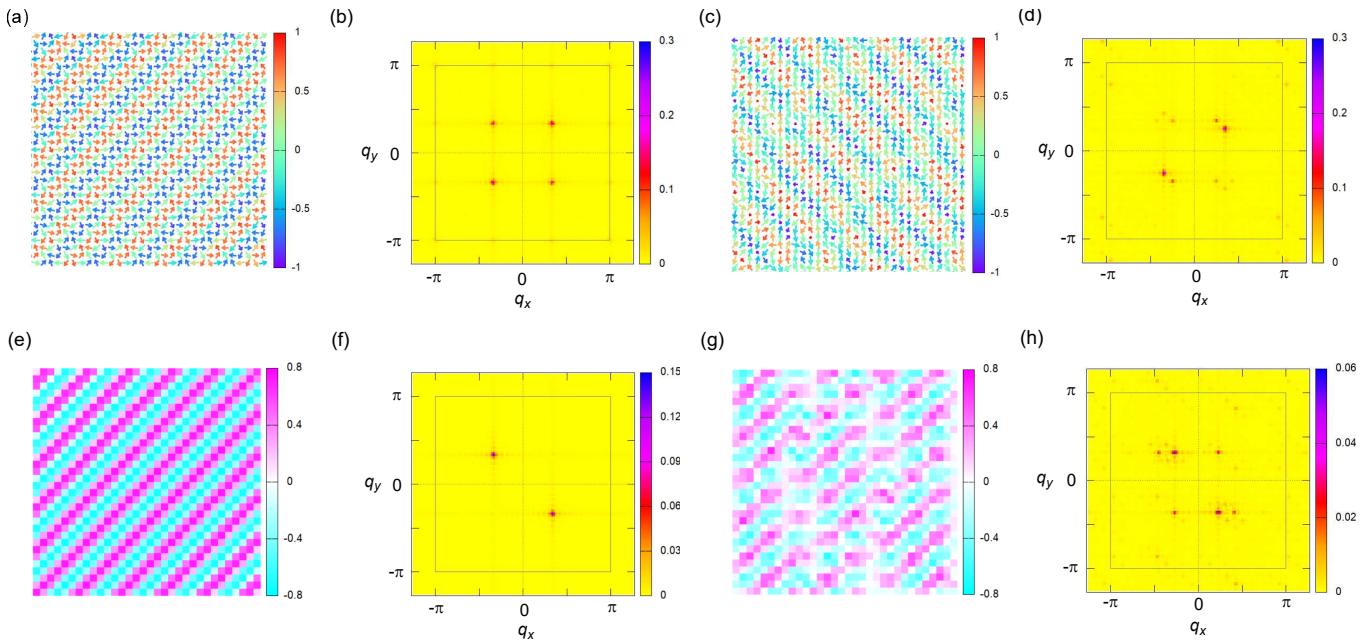}
\caption{The results of the KPM-LD simulations for $J=0.1$: (a,b,e,f) without the SOC ($\alpha=\beta=0$) and (c,d,g,h) with the Rashba SOC ($\alpha=0.2$ and $\beta=0$). 
(a) and (b) represent a typical spin pattern and the norm of its Fourier transform, $|{\mathbf S}_{\mathbf q}|$, of the $2Q$-vortex state, while (c) and (d) correspond to those of the multiple-$Q$ state.
(e) and (f) are the real-space pattern of the spin scalar chirality and the absolute value of its Fourier transform, $|\kappa_{\mathbf q}|$, for the $2Q$-vortex state in (a).
(g) and (h) display those of the multiple-$Q$ state in (c).
The real-space textures of spin and scalar chirality in (a), (c), (e), and (g) are shown for a part of the whole system with $N=96^2$ system for clarity.
In (a) and (c), the arrows denote the directions of the localized spins in the $xy$ plane, and the color represents the $z$ component.
}
\label{fig:Rashba_spin}
\end{figure*}

Next, we discuss the magnetic orderings in the presence of only Rashba SOC ($\alpha\neq0$ and $\beta=0$).
First of all, we show the Fermi surface instabilities.
In this section we introduce the TNN terms with $t_3=-0.5$ and $\alpha_3=-0.5\alpha$, and set the chemical potential as $\mu=0.98$, following the previous study on the SOC-free case~\cite{Ozawa2016}.
Figures~\ref{fig:Rashba_Fermisurface}(a) and \ref{fig:Rashba_Fermisurface}(b) show the Fermi surface and the bare magnetic susceptibility in the absence of SOC.
The Fermi surfaces show rather strong partial nesting with ${\mathbf Q}_1=(\pi/3,\pi/3)$ and ${\mathbf Q}_2=(-\pi/3,\pi/3)$ as shown in Fig.~\ref{fig:Rashba_Fermisurface}(a), which leads to the distinct peaks in the susceptibility at the same wave vectors as shown in Fig.~\ref{fig:Rashba_Fermisurface}(b).

When the Rashba SOC is introduced, the spin degeneracy of the Fermi surface is lifted and accordingly the peaks in the susceptibility are split in a complicated way. 
Figures~\ref{fig:Rashba_Fermisurface}(c)-\ref{fig:Rashba_Fermisurface}(f) display the spin-split Fermi surfaces and the largest eigenvalues of the bare magnetic susceptibility ${\lambda}^1_{\mathbf q}$ for $\alpha=0.2$ and $\beta=0$.
Due to the spin-splitting of the Fermi surface, the two peaks in the SOC-free susceptibility at ${\mathbf Q}_1$ and ${\mathbf Q}_2$ split into totally fourteen distinct peaks with almost equal amplitudes [Figs.~\ref{fig:Rashba_Fermisurface}(d) and \ref{fig:Rashba_Fermisurface}(f)].
As shown in Fig.~\ref{fig:Rashba_Fermisurface}(f) we denote these wave vectors as ${\mathbf Q}^\eta_\nu$ with $\nu=1-4$ for $\eta=a,b,c$ and $\nu=1,2$ for $\eta=d$. 
Note that a set of wave vectors indexed with a superscript $\eta$, $\left\{{\mathbf Q}^\eta_\nu\right\}$, are related with each other by $C_4$ and $\sigma_v$ symmetry operations, yielding the exactly identical value of $\lambda^1_{\mathbf q}$. 
As displayed in Figs.~\ref{fig:Rashba_Fermisurface}(c) and \ref{fig:Rashba_Fermisurface}(e) we can assign $\left\{{\mathbf Q}^a_\nu\right\}$ ($\left\{{\mathbf Q}^b_\nu\right\}$) to the wave vectors connecting two portions within the outer (inner) Fermi surfaces, while $\left\{{\mathbf Q}^c_\nu\right\}$ ($\left\{{\mathbf Q}^d_\nu\right\}$) to the wave vectors connecting from one in the inner (outer) Fermi surface to the other in the outer (inner). 
After closely comparing the competing heights of those peaks for large system sizes, we find out that ${\lambda}^1_{\mathbf q}$ at $\left\{{\mathbf Q}^a_\nu\right\}$ are slightly larger than the others [Figs.~\ref{fig:Rashba_Fermisurface}(c) and \ref{fig:Rashba_Fermisurface}(d)]. 

In the following we discuss the results of the KPM-LD simulations.
Figures~\ref{fig:Rashba_spin} show the simulation results for $J=0.1$.
As already reported in details in the previous study~\cite{Ozawa2016}, without the SOC a noncoplanar double-$Q$ ordering appears, characterized with ${\mathbf Q}_1$ and ${\mathbf Q}_2$ [Figs.~\ref{fig:Rashba_spin}(a) and \ref{fig:Rashba_spin}(b)]. 
The double-$Q$ ordering, named $2Q$-vortex, shows a stripe of the spin scalar chirality [Figs.~\ref{fig:Rashba_spin}(e) and \ref{fig:Rashba_spin}(f)]~\cite{Ozawa2016}. 

With the introduction of the Rashba SOC with $\alpha=0.2$, we find a more complex multiple-$Q$ state, as shown in Figs.~\ref{fig:Rashba_spin}(c) and \ref{fig:Rashba_spin}(d).
As stated in Sec.~\ref{sec:KPMLD}, for $\alpha=0.2$ we performed the KPM-LD by adopting several different spin ansatzes as the initial spin configurations, which are double- or single-$Q$ orderings constructed from $\left\{{\mathbf Q}^a_\nu\right\}$ [see Eq.~(\ref{eq:rashba_ansatz})].
In Figs.~\ref{fig:Rashba_spin}(c) and \ref{fig:Rashba_spin}(d) we only show the results obtained from one of the initial ansatzes. We stress that for the other ansatzes we have also confirmed similar multiple-$Q$ orderings with the same energy within the resolution of the KPM-LD, which are characterized with the same set of wave vectors as in Fig.~\ref{fig:Rashba_spin}(d), although the weight distributions among them vary to some extent.
As seen in Fig.~\ref{fig:Rashba_spin}(d), the multiple-$Q$ ordering appears to be dominantly formed by $\left\{{\mathbf Q}^a_\nu\right\}$ as well as other closely-located wave vectors such as ${\mathbf Q}^c_3$ and ${\mathbf Q}^c_4$ [see Fig.~\ref{fig:Rashba_Fermisurface}(f)]. 

We find that the multiple-$Q$ order exhibits a modulated stripe of the spin scalar chirality whose net component vanishes, as shown in Figs.~\ref{fig:Rashba_spin}(g) and \ref{fig:Rashba_spin}(h).
The spatial pattern of the spin scalar chirality makes the multiple-$Q$ order distinct from SkXs, which, in general, show a nonzero net value of the scalar chirality.
The discovery of such a complex multiple-$Q$ ordering is remarkable as compared with localized or continuum spin models with only NN interactions under the same symmetry, in which simpler multiple-$Q$ orderings like SkX are normally found~\cite{Banerjee2014,Li2014}.
Although the effective spin model describing the RKKY interaction also predicts the stabilization of multiple-$Q$ orderings~\cite{Hayami2018}, our study on the original Kondo lattice model indicates potential fomation of further complex multiple-$Q$ orderings characterized with more than two wave vectors, which would be attributed to the full integration of the conduction electrons to the simulations. 

\begin{figure}[!htb]
\begin{center}
\includegraphics[width=\linewidth,clip]{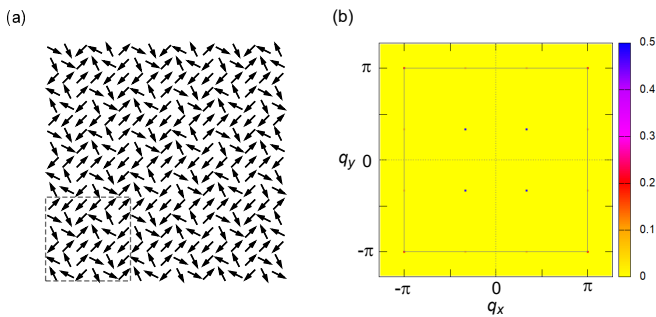}
\end{center}
\caption{(a) Spin texture and (b) $|{\mathbf S}_{\mathbf q}|$ of the coplanar $2Q$-flux order, constructed from Eq. (\ref{eq:flux}). 
The dashed square in (a) denotes the manetic unit cell.}
\label{fig:flux}
\end{figure}

\begin{figure}[!htb]
\begin{center}
\includegraphics[width=0.8\linewidth,clip]{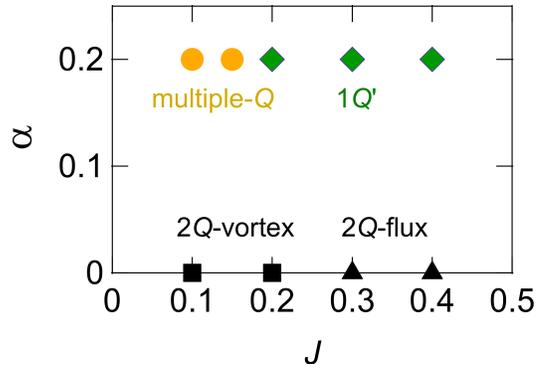}
\end{center}
\caption{Ground-state phase diagram for the Kondo lattice model with the Rashba SOC ($\alpha\neq0$ and $\beta=0$), determined by the KPM-LD simulations.}
\label{fig:Rashba_phase_diagram}
\end{figure}

Then we discuss the evolution of the magnetic orderings while increasing $J$.
The KPM-LD simulations reveal that without the SOC the noncoplanar $2Q$-vortex state is favored for $J=0.1$ and $0.2$, while it is replaced by a coplanar double-$Q$ ordering for $J=0.3$ and $0.4$, as represented in Fig. \ref{fig:flux}. 
We refer to the latter ordering as $2Q$-flux. 
For $\alpha=0.2$ the complex multiple-$Q$ orders are stabilized for $J=0.1$ and $0.15$, whereas a single-$Q$ ordering is favored for $J=0.2-0.4$. 
We note that the ordering vector of the single-$Q$ ordering found for large $J$, named $1Q'$, is not any of $\left\{{\mathbf Q}^\eta_\nu\right\}$ but another relatively large wave vector denoted as the white arrows in Fig. \ref{fig:Rashba_Fermisurface}(d), around which the susceptibility takes a broad peak with a sizable height.
We summarize the results in the $J$-$\alpha$ phase diagram in Fig.~\ref{fig:Rashba_phase_diagram}.

\begin{figure}[!htb]
\begin{center}
\includegraphics[width=0.8\linewidth,clip]{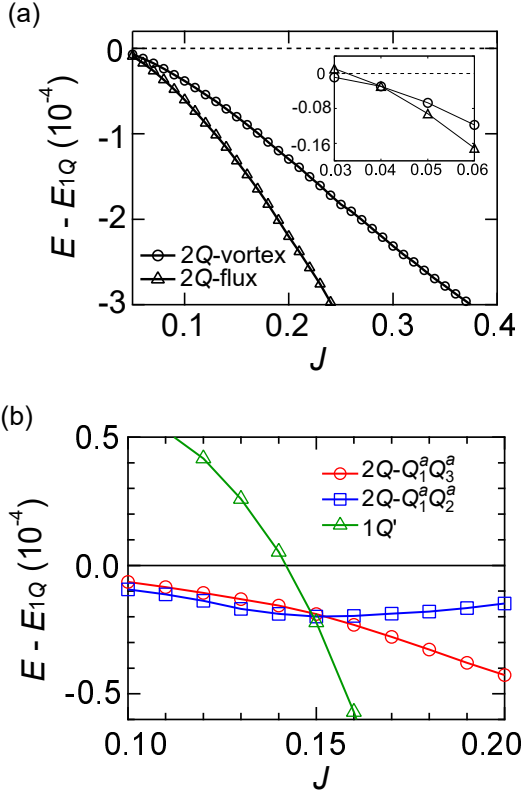}
\end{center}
\caption{
$J$ dependence of the energies for several ansatzes, estimated by variational calculations with (a) $\alpha=\beta=0$ and (b) $\alpha=0.2$ and $\beta=0$.
(a) represents the energies for $2Q$-vortex and $2Q$-flux, measured from that of the $1Q$ helical ordering.
The inset of (a) is the same plot in the small $J$ region. 
(b) shows the energies for two ansatzes for $2Q$ states and $1Q$', measured from that of the $1Q$ helical ordering.
The calculations are done for $N=480^2$.
}
\label{fig:Rashba_variational}
\end{figure}

Complementary to the KPM-LD we perform variational calculations. 
In the absence of the Rashba SOC we compare the energies of $2Q$-vortex and $2Q$-flux in Fig. \ref{fig:Rashba_variational}(a). 
$2Q$-vortex is described as~\cite{Ozawa2016} 
\begin{equation} 
{\mathbf S}_{j} = 
\left(
    \begin{array}{ccc}   
     \sqrt{1-b^2\sin^2(\mathbf{Q}_2\cdot\mathbf{r}_j)}\cos(\mathbf{Q}_1\cdot\mathbf{r}_j)\\
     \sqrt{1-b^2\sin^2(\mathbf{Q}_2\cdot\mathbf{r}_j)}\sin(\mathbf{Q}_1\cdot\mathbf{r}_j)\\
     b\sin(\mathbf{Q}_2\cdot\mathbf{r}_j)\\
    \end{array}
  \right),
\label{eq:vortex}
\end{equation}
while $2Q$-flux is found to be represented as
\begin{equation} 
{\mathbf S}_{j} = \hat{N}\left(
    \begin{array}{ccc}  
      \cos(\mathbf{Q}_1\cdot\mathbf{r}_j)\\
      \cos(\mathbf{Q}_2\cdot\mathbf{r}_j)\\
      0\\
    \end{array}
  \right).
\label{eq:flux}
\end{equation}
In Fig. \ref{fig:Rashba_variational}(a) we set the variational parameter $b$, which describes the noncoplanarity, at $b=0.6$ for the $2Q$-vortex ansatz in Eq. (\ref{eq:vortex}). Note that in Fig. \ref{fig:Rashba_variational}(a) we subtract the energy of the single-$Q$ helical ordering corresponding to the $b=0$ case in Eq. (\ref{eq:vortex}).
Figure \ref{fig:Rashba_variational}(a) shows that for $J\lesssim0.04$, $2Q$-vortex has lower energy than $2Q$-flux and vice versa for $J\gtrsim0.04$. 
We also see that the helical ordering is unfavored in the whole range of $J$ studied here.
Thus, the variational calculations verify the trend in the KPM-LD that $2Q$-vortex transitions to $2Q$-flux while increasing $J$.
The critical value of $J$ is considerably different between the two calculations, which might be attributed to the energy resolution of the KPM-LD or the incompleteness of the variational ansatzes. 

Figure \ref{fig:Rashba_variational}(b) shows the energy comparison among several ansatzes for $\alpha=0.2$ and $\beta=0$.
Since the multiple-$Q$ states discovered in the KPM-LD, e.g., Figs. \ref{fig:Rashba_spin}(c) and \ref{fig:Rashba_spin}(d), are too complicated to deduce the corresponding ansatzes, we simply employ the double-$Q$ orderings that are used for the initial spin configurations in the KPM-LD [see Eq. (\ref{eq:rashba_ansatz})], which maximize the energy gain of the generalized RKKY Hamiltonian in Eq. (\ref{eq:rkky_diag}) without the normalization factor.
In Fig. \ref{fig:Rashba_variational}(b) we denote the double-$Q$ orderings formed by ${\mathbf Q}^a_{\nu_1}$ and ${\mathbf Q}^a_{\nu_2}$ as $2Q$-$Q^a_{\nu_1}Q^a_{\nu_2}$. 
Here the energies are measured from that of the single-$Q$ ordering formed by ${\mathbf Q}^a_\nu$, named $1Q$.
Although the ansatzes for the multiple-$Q$ states are approximate ones, it turns out that they qualitatively reproduce the $J$ dependence obtained by the KPM-LD shown in Fig.~\ref{fig:Rashba_phase_diagram}: the double-$Q$ orderings are favored up to $J\sim0.15$ and replaced by $1Q$' for $J\gtrsim0.15$.

\subsection{Case with $\alpha=0$ and $\beta\neq0$}\label{sec:Dresselhaus}
Finally we discuss the case with the Dresselhaus SOC only ($\alpha=0$ and $\beta\neq0$).
As stated in Sec.~\ref{sec:exchange} magnetic orders for the Dresselhaus-only case are identical to what are obtained by a $\pi$ rotation of those for the Rashba-only case along the [110] axis [see Eq.~(\ref{eq:exchangeRD_spintransform})].
Hence the phase diagram for the Rashba-only case presented in Fig.~\ref{fig:Rashba_phase_diagram} is common to the Dresselhaus-only case, with the simple global rotation applied to the magnetic orders. 

\begin{figure}[!htb]
\centering
\includegraphics[width=\columnwidth,clip]{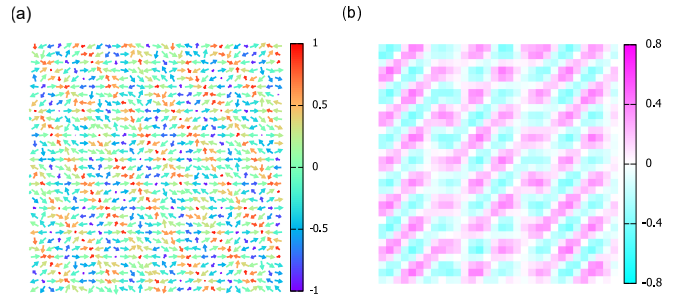}
\caption{(a) Multiple-$Q$ order with $J=0.1$ for $\alpha=0$ and $\beta=0.2$. This is produced by applying the $\pi$ rotation along the [110] axis to the spin texture for $\alpha=0.2$ and $\beta=0$ shown in Fig.~\ref{fig:Rashba_spin}(c).
(b) Spin scalar chirality of the multiple-$Q$ order in (a), which is identical to the Rashba-only case shown in Fig.~\ref{fig:Rashba_spin}(g). 
}
\label{fig:Dresselhaus_spin}
\end{figure}

Figure~\ref{fig:Dresselhaus_spin}(a) shows the multiple-$Q$ order in the Dresselhaus-only case with $J=0.1$, which is obtained by applying the $\pi$ rotation to the one for the Rashba-only case in Fig.~\ref{fig:Rashba_spin}(c).
The uniform rotation leads to the same spatial pattern of the spin scalar chirality as the Rashba-only case, as shown in Fig.~\ref{fig:Dresselhaus_spin}(b).
We also remark on the distinction of the multiple-$Q$ ordering here from those expected in localized spin models describing only NN interactions with the same $D_{2d}$ symemtry; in the latter case shows up a periodic array of antiskyrmions, which are characterized with the opposite sign of the topological invariant to conventional Skyrmions~\cite{Nayak2017}. 

\section{Conclusions}\label{sec:concl}
To summarize, we have studied magnetic orderings generated by itinerant electrons subject to the Rashba ($\alpha$) and Dresselhaus ($\beta$) SOCs by means of the large-scale numerical simulations as well as the variational calculations based on the perturbation analyses.
We discovered the complex multiple-$Q$ orderings under zero magnetic field, depending on the nature of the spin-split Fermi surfaces induced by the SOCs. 
For the equal strength of both SOCs ($\alpha=\beta\neq0$) the exotic spin texture is unveiled in a broad range of $J$, characterized with as many as six wave vectors.
Notably this sextuple-$Q$ ordering shows a checkerboard-like pattern of the spin scalar chirality.
In the case that only Rashba or Dresselhaus SOC exists ($\alpha$ or $\beta=0$) we found another type of complex multiple-$Q$ states, which are distinct from those expected in localized spin systems under the same symmetry. 
Our findings suggest that the combination of the spin-charge and spin-orbit couplings under broken spatial inversion symmetry gives rise to richer multiple-$Q$ magnetic orders than the competition between the ferromagnetic and DM interactions in localized spin systems. 
Our theory would be potentially applicable to noncentrosymmetric $f$-electron compounds as well as heterostuctures of spin-orbit coupled metals and magnetic materials. 

\begin{acknowledgments}
We thank K. Barros and R. Ozawa for providing us with the code for the KPM-LD simulations.
We are also grateful to R. Ozawa and S. Iino for fruitful discussions.
K.N.O. acknowledges Y. Tserkovnyak for his incisive comments.
The KPM-LD simulations were carried out at the Supercomputer Center, Institute for Solid State Physics, University of Tokyo.
K.N.O. is supported by the Japan Society for the Promotion of Science through a research fellowship for young scientists.
\end{acknowledgments}

\bibliography{citation} 

\end{document}